\documentstyle[preprint,aps]{revtex}

\begin{document}
\title{The Interference Term between the Spin and Orbital Contributions to
$M1$ Transitions}
\author{M.S. Fayache$^1$, L. Zamick$^2$, Y.Y. Sharon$^2$ and Y. Durga 
Devi$^2$\\
(1) D\'{e}partement de Physique, Facult\'{e} des Sciences de Tunis\\
Tunis 1060, Tunisia\\
\noindent (2) Department of Physics and Astronomy, Rutgers University\\
Piscataway, New Jersey 08855\\}
\date{\today}
\maketitle
\begin{abstract}
We study the cross-correlation between the spin and orbital parts of 
magnetic dipole transitions $M1$ in both isoscalar and isovector channels.
In particular, we closely examine certain cases where $\sum B(M1)$ is very 
close to $\sum B(M1)_{\sigma}$ + $\sum B(M1)_l$, implying a cancellation of 
the summed interference terms. We gain some insight into this problem by 
considering special cases approaching the $SU(3)$ limit, and by examining 
the behaviour of single-particle transitions at the beginning and towards 
the end of the $s-d$ shell.
\end{abstract}

\section{Introduction}
In a previous work \cite{fay1}, the magnetic dipole transitions from the
ground states to excited states of several nuclei in the $s-d$ shell were
calculated: $^{20}Ne$, $^{22}Ne$, $^{24}Mg$, $^{28}Si$, $^{32}S$, and 
$^{36}Ar$. The main focus of the work was on how the transition strengths were 
affected by the strengths of the spin-orbit and tensor interactions inside 
a nucleus, and near the end of the paper, it was briefly mentioned that a new 
topic would be of interest: ``the cross-correlation between the spin and 
orbital parts of $B(M1)$''. In the present work, we wish to elaborate on this 
point. 

As in the previous work, we define the $M1$ transition strength as: 

\begin{equation}
B(M1)\! \uparrow~=~\frac{1}{(2J_i+1)}\sum_{M_f,\mu, M_i}|\langle
\Psi_{M_f}^{J_f}~A_{\mu}~\Psi_{M_i}^{J_i}\rangle |^2 \;,
\label{eq:bm1}
\end{equation}

where

\begin{equation}
\vec{A}=\sqrt{\frac{3}{4\pi}} \left [
\sum_{i=1}^Z (g_{l\pi}\vec{l_i}+g_{s\pi}\vec{s_i}) +
\sum_{j=1}^N(g_{l\nu}\vec{l_j}+g_{s\nu}\vec{s_j})\right] \:.
\label{eq:avector}
\end{equation}

We define three $B(M1)$'s: total, spin, and orbital with the following 
parameters (in units of $\mu_N$)

\begin{tabbing}
Total $\;\;\;B(M1)$:~~\= $g_{l\pi}=1$, \= $g_{s\pi}=5.586$, \= $g_{l\nu}=0$, \=
$g_{s\nu}= -3.826$\\
Orbital $B(M1)_l$:~~\> $g_{l\pi}=1$, \> $g_{s\pi}=0$, \> $g_{l\nu}=0$, \> $g_{s\nu}=0$\\\
Spin $B(M1)$:~~\> $g_{l\pi}=0$, \> $g_{s\pi}=5.586$, \> $g_{l\nu}=
0$,\> $g_{s\nu}= -3.826$\\
\end{tabbing}

As the title of this work implies, we wish to study the interference terms 
between the spin and orbital parts. For a transition to an individual state, 
we can write 

\begin{eqnarray*}
B(M1) & = & (\sqrt{B(M1)_l} \pm \sqrt{B(M1)_{\sigma}})^2\\
      & = & B(M1)_l + B(M1)_{\sigma} \pm 2\sqrt{B(M1)_l}\sqrt{B(M1)_{\sigma}}
\nonumber\\
\end{eqnarray*}

For some states, we would have a plus sign (constructive interference), and 
for some the minus sign (destructive interference). We can also consider the 
summed strength to all calculated states. It was already noted in our previous 
work \cite{fay1} that for some cases ``$\sum B(M1)$ is very close to 
$\sum B(M1)_{\sigma}$ + $\sum B(M1)_l$''. This would imply a cancellation of 
the summed interference terms. In this work, we will study this in a more 
quantitative manner.

\section{Calculations}

The interaction that was used was described in \cite{fay1}, so we will be 
brief. We used the ($x,y$) interaction 

\begin{equation}
V(r)=V_c(r)+x \cdot V_{s.o.}+y \cdot V_t
\label{eq:xyint}
\end{equation}

\noindent where $s.o.$ stands for the two-body spin-orbit interaction, $t$
for the tensor interaction, and $V_c(r)$ is everything else, especially
the (spin-dependent) central interaction. We can vary the strength of the 
spin-orbit and tensor interactions by varying $x$ and $y$. The optimum fit 
to a free $G$-matrix is obtained with $x=1,~y=1$ \cite{zheng}. Arguments could 
be made that these parameters should be changed inside a nucleus.

As an example, we show results for $M1$ transitions in $^{28}Si$ with the 
($x,y$) interaction for $x=1,~y=1$. In Table I we show results for isoscalar 
transitions from the ground state ($T=0,~J=0^+$) to excited states 
($T=0,~J=1^+$) in units of $10^{-2}\mu_N^2$. In Table II we give results 
for isovector transitions in $^{28}Si$ in units of $\mu_N^2$. We show 
only the first ten states, but the sum $\sum B(M1)$ is over all states 
(around 500). We also show the sign of the interference term.

In Tables III and IV we show respectively the isoscalar and isovector summed 
$M1$ strengths for the ($x,y$) interaction with $x=1,~y=1$. We also show the 
deviation $\Delta$ which is equal to $\sum B(M1)- \sum B(M1)_{\sigma} - 
\sum B(M1)_l$. We also introduce an angle $\theta$ in order to better 
describe the interference:

\begin{equation} \sum B(M1)= \sum B(M1)_{\sigma} + \sum B(M1)_l + 
2\sqrt{\sum B(M1)_l} \sqrt{\sum B(M1)_{\sigma}}cos(\theta) 
\end{equation}

\noindent In order for the interference term to vanish, we must have $\theta=
90^{\circ}$. 

By examining Tables III and IV, we see that there is a striking contrast 
between the isoscalar and isovector cases. In the former case, we have 
$\theta = 180^{\circ}$ for all cases. For the isovector case, $\theta$ is 
closer to 90$^{\circ}$. When $\theta$ is exactly 90$^{\circ}$, the sum of the 
interference terms over all states would vanish.

The isoscalar result is easy to understand. Consider the total angular 
momentum operator $\vec{J}=\vec{L}+\vec{S}$. Clearly the matrix element 
$\langle 1^+ \vec{J}~0^+ \rangle$ will vanish. Hence $\langle 1^+ \vec{L}~ 0^+ 
\rangle$=~-$\langle 1^+ \vec{S}~ 0^+ \rangle$ for each $1^+$ state. Thus, if 
the isoscalar $M1$ operator is written as $a\vec{L}+b\vec{S}$, then a 
transition matrix element from the ground state to a state $1^+_i~(T=0)$ is 
given by:

\begin{equation}
\langle 1^+_i~a\vec{L}+b\vec{S}~ 0^+ \rangle=(a-b)\langle 1^+_i \vec{L}~ 0^+
\rangle
\end{equation}

\noindent The summed $B(M1)$ strength is $(a-b)^2\sum_i
|\langle 1^+_i \vec{L}~ 0^+\rangle|^2$, and as long as $a$ and $b$ have the 
same sign, it is easy to see that $\theta$ is equal to $180^{\circ}$. Indeed, 
$a=(g_{l\pi}+g_{l\nu})/2=0.5$ and $b=(g_{s\pi}+g_{s\nu})/2=0.88$, and they do 
have the same sign.

For the isovector case, however, we do not have such a constraint, so that the
 signs of the interference terms are more random, and the angle $\theta$ is 
closer to $90^{\circ}$. We cannot help but notice that the deviation from 
$90^{\circ}$ increases as we increase the mass number in the $1s-0d$ shell. 
For $^{20}Ne$, $\theta$ is equal to $90.00^{\circ}$, but it increases steadily 
to $90.45^{\circ}$, $92.31^{\circ}$, $93.33^{\circ}$ and $95.45^{\circ}$ for 
$^{24}Mg$, $^{28}Si$, $^{32}S$, $^{36}Ar$. This suggests that $\theta$
gets closer to $90^{\circ}$ as we get closer to Wigner's U(4) limit
\cite{Wigner}, which includes the $SU(3)$ limit of Elliott \cite{Elliott}. The 
$SU(3)$ model holds much better in the lower half of the $s-d$ shell than in 
the upper half. In the extreme $SU(3)$ limit, the spin $M1$'s will vanish, and 
in that case $B(M1)= B(M1)_l$. The interference term will vanish trivially. 
We can see from Table II, however, that for $^{28}Si$ the summed isovector 
spin and orbital strengths are almost the same, so that it is a non-trivial 
result that $\theta$ is close to $90^{\circ}$ in this case. 

Another way of looking at this is to note that the results get better as the 
spin-orbit interaction strength is decreased. 

In Table V we consider two cases where $N$ does not equal $Z$: $^{10}Be$ and 
$^{22}Ne$. We break up the contributions into two parts: $J=0^+~T=1 
\rightarrow J=1^+~T=1$ and $J=0^+~T=1 \rightarrow J=1^+~T=2$. We then consider
 the total contribution. For both nuclei we see large deviations from 
$90^{\circ}$. For example, in $^{22}Ne$ the value of $\theta$ for $T=1 
\rightarrow T=1$ is 85.965$^{\circ}$, whilst for $T=1 \rightarrow T=2$ it is 
102.488$^{\circ}$. The combined result yields 88.748$^{\circ}$. What this 
tells us is that when we consider all transitions, it appears that we are 
close to randomness of the sign of the interference term. However, if we 
consider each part separately, there is a large deviation from randomness. 
For $T=1 \rightarrow T=1$ the signs conspire so as to enhance the total 
$B(M1)$, whilst for $T=1 \rightarrow T=2$ they act to diminish the total 
$B(M1)$. In combination, the two effects oppose each other and yield a result 
close to randomness of the phase.

For a transition from $j=l+1/2$ to $j=l+1/2$ ($e.g.$ $d_{5/2} 
\rightarrow d_{5/2}$), the isovector matrix element is proportional to 
$(l+\mu_p-\mu_n)=(l+4.766)$. Thus the interference term will be positive 
corresponding to $\theta=0^{\circ}$. This should be relevant to the lower 
part of the $s-d$ shell. 
For the transition $j=l-1/2$ to $j=l-1/2$ ($e.g.$ $d_{3/2} \rightarrow 
d_{3/2}$), the isovector matrix element is proportional to 
$[l+1-(\mu_p-\mu_n)]=(l+1~-4.706)$. In this case, the interference term will 
be negative, corresponding to $\theta=180^{\circ}$. This will be most 
important near the end of the $s-d$ shell $e.g.$ for $^{36}Ar$. The matrix 
element $j=l \pm 1/2 \rightarrow j=l \mp 1/2$ is proportional to 
$(g_l-g_s)=[l-2(\mu_p - \mu_n)]$. Here again the interference term is 
negative, corresponding to $\theta=180^{\circ}$. 

In Table VI we vary the parameters of our $(x,y)$ interaction in order to see 
how the `interference angle $\theta$' depends on the spin-orbit and tensor 
interactions. We use $^{28}Si$ as an example. We see that the smaller the 
value of $x$ (the strength of the spin-orbit interaction), the closer $\theta$ 
is to $90^{\circ}$. For $y=0$ (no tensor interaction present), the values of 
$\theta$ for $x=0$ and $x=1$ are respectively $90.03^{\circ}$ and 
$94.93^{\circ}$. With $y=1$ (full strength interaction), the corresponding 
values are $89.93^{\circ}$ and $92.33^{\circ}$. This behaviour is consistent 
with the well-known fact that the spin-orbit interaction destroys the $SU(3)$ 
symmetry. Concerning the tensor interaction, the behaviour can be explained 
by the fact that in an open-shell nucleus this interaction behaves somewhat 
like a spin-orbit interaction but with sign opposite to that of the basic 
spin-orbit interaction \cite{wong}. Thus, for $x=1~y=1$, the value of 
$\theta$ is smaller than for $x=1~y=0$. The values are $92.33^{\circ}$ and 
$94.93^{\circ}$ respectively. The effective spin-orbit interaction for 
$x=1~y=1$ is weaker than for $x=1~y=0$, and so we are closer to the $SU(3)$ 
limit. 

Whereas in free space the choice $x=1~y=1$ gives the best results in 
comparison with $G$ matrices obtained from realistic interactions, there is 
some evidence discussed in \cite{fay1} that inside a nucleus the spin-orbit 
interaction should be stronger than in free space, and the tensor interaction 
weaker. We therefore also consider the case $x=1.5~y=0.5$. Because the 
spin-orbit interaction is stronger, we find that the interference angle 
deviates from $90^{\circ}$ $i.e.$ $\theta=98.31^{\circ}$ in this case.

\section{Closing Remarks}

When we are close to the $U(4)$ limit of Wigner \cite{Wigner}, we find that for
isovector transitions in $N=Z$ nuclei the sum $\sum B(M1)$ is close to 
$\sum B(M1)_{\sigma} + \sum B(M1)_l$. More generally, we have defined an 
interference angle $\theta$ in Eq. (4). For $x=0$ ($i.e.$ no spin-orbit 
interaction present), $\theta$ is very close to $90^{\circ}$ and the sum of 
all the interference terms is almost zero (randomness). As we increase the 
spin-orbit splitting by increasing $x$, the angle $\theta$ becomes larger than 
$90^{\circ}$. Also, in nuclei where $SU(3)$ symmetry is not so good, 
$\theta$ becomes larger than $90^{\circ}$. For isoscalar transitions in $N=Z$ 
nuclei, we find that we have maximum destructive interference between the 
orbital and spin $M1$ amplitudes. This can be explained by the fact that the 
total angular momentum operator cannot induce $M1$ transitions. For $N \neq Z$ 
nuclei, like $^{22}Ne$, and if we consider all transitions, it looks like we 
are close to randomness ($\theta=88.7^{\circ}$). However, if we look at each 
final isospin separately, the picture is changed $e.g.$: for 
$T=1 \rightarrow T=1$ transitions $\theta$ is $85.9^{\circ}$ (net constructive 
interference), whilst for $T=1 \rightarrow T=2$ $\theta$ is $102.7^{\circ}$ 
(considerable destructive interference).

This work was supported in part by a D.O.E. grant DE-FG02-95ER-40940.

\nopagebreak

\begin{table}
\caption{$B(M1)$ Transitions in $^{28}Si$: Isoscalar Case 
$T=0 \rightarrow T=0$ in units of $10^{-2}\mu_N^2$. The sign of the 
interference term is shown.}
\begin{tabular}{ccccc}
Energy ($MeV$) & $B(M1)$ & $B(M1)_l$ & $B(M1)_{\sigma}$ & Sign\\
\tableline
 9.394 & 0.4514 & 0.7814 & 2.4210 & -\\
 9.817 & 0.1440 & 0.2493 & 0.7724 & -\\
10.424 & 0.0428 & 0.0742 & 0.2298 & -\\
12.237 & 0.0831 & 0.1438 & 0.4456 & -\\
12.709 & 0.2088 & 0.3615 & 1.1200 & -\\
13.350 & 0.1011 & 0.1749 & 0.5419 & -\\
13.557 & 0.0000 & 0.0001 & 0.0001 & -\\
13.706 & 0.0019 & 0.0033 & 0.0101 & -\\
14.219 & 0.0061 & 0.0106 & 0.0328 & -\\
14.497 & 0.0133 & 0.0231 & 0.0715 & -\\
\tableline
$\sum B(M1)$ & 1.6876 & 2.9216 & 9.0511 & -\\
\end{tabular}
\end{table}

\begin{table}
\caption{$B(M1)$ Transitions in $^{28}Si$: Isovector Case 
$T=0 \rightarrow T=1$ in units of $\mu_N^2$.}
\begin{tabular}{ccccc}
Energy ($MeV$) & $B(M1)$ & $B(M1)_l$ & $B(M1)_{\sigma}$ & Sign\\
\tableline
10.310 & 1.3520 & 0.0552 & 0.8603 & +\\
11.768 & 0.4906 & 0.1224 & 0.1229 & +\\
12.660 & 0.0344 & 0.1761 & 0.5490 & -\\
12.970 & 0.3874 & 0.3591 & 0.0005 & +\\
13.318 & 0.0612 & 0.0296 & 0.0057 & +\\
13.622 & 0.0339 & 0.0769 & 0.0087 & -\\
13.901 & 0.0111 & 0.0137 & 0.0494 & -\\
14.050 & 0.0544 & 0.0670 & 0.0007 & -\\
14.535 & 0.3469 & 0.3989 & 0.0018 & -\\
15.044 & 0.1597 & 0.0013 & 0.1900 & -\\
\tableline
$\sum B(M1)$ & 4.4449 & 2.2424 & 2.3911 & \\
\end{tabular}
\end{table}

\begin{table}
\caption{Summed $B(M1)$ Strength: Isoscalar Case
$T=0 \rightarrow T=0$ in units of $10^{-2}\mu_N^2$.}
\begin{tabular}{cccccc}
Nucleus & $\sum B(M1)$ & $\sum B(M1)_l$ & $\sum B(M1)_{\sigma}$ & 
$\Delta$\tablenotemark[1] & $\theta$\tablenotemark[2]\\
\tableline
$^{20}Ne$ & 0.3925 & 0.6794 & 2.1047 & -2.3916 & 180$^{\circ}$\\
          &        &        &        &         & \\
$^{24}Mg$ & 1.1810 & 2.0451 & 6.3341 & -7.1982 & 180$^{\circ}$\\
          &        &        &        &         & \\
$^{28}Si$ & 1.6876 & 2.9216 & 9.0611 & -10.2851 & 180$^{\circ}$\\
          &        &        &        &         & \\
$^{32}S$  & 1.9131 & 3.3137 & 10.2640 & -11.6646 & 180$^{\circ}$\\
          &        &        &        &         & \\
$^{36}Ar$ & 1.5189 & 2.6299 & 8.1467 & -9.2577 &  180$^{\circ}$\\
\end{tabular}
\tablenotetext[1]{$\Delta=\sum B(M1)- \sum B(M1)_{\sigma} - 
\sum B(M1)_l$}
\tablenotetext[2]{$cos(\theta)=\Delta/(2\sqrt{(\sum B(M1)_l)
(\sum B(M1)_{\sigma})})$}
\end{table}

\begin{table}
\caption{Summed $B(M1)$ Strength: Isovector Case
$T=0 \rightarrow T=1$ in units of $\mu_N^2$.}
\begin{tabular}{cccccc}
Nucleus & $\sum B(M1)\tablenotemark[1]$ & $\sum B(M1)_l$ & $\sum B(M1)_{\sigma}$ & 
$\Delta$ & $\theta(^{\circ})$\\
\tableline
$^8Be$    & 1.055  & 0.6701 & 0.3784 & 0.0065  & 89.60\\
          &        &        &        &         & \\
$^{20}Ne$ & 1.5326 & 0.9456 & 0.5817 & -0.00002 & 90.00\\
          &        &        &        &         & \\
$^{24}Mg$ & 3.7797 & 1.7302 & 2.0793 & -0.0298 & 90.40\\
          &        &        &        &         & \\
$^{28}Si$ & 4.4449 & 2.2424 & 2.3911 & -0.1886 & 92.30\\
          &        &        &        &         & \\
$^{32}S$  & 4.7569 & 2.7134 & 2.3181 & -0.2926 & 93.33\\
          &        &        &        &         & \\
$^{36}Ar$ & 3.3485 & 2.2480 & 1.4426 & -0.3421 & 95.40\\
\end{tabular}
\tablenotetext[1]{In this table, we use the interaction given in Eq. (3), with 
$x=1~y=1$.}
\end{table}

\begin{table}
\caption{Summed $B(M1)$ Strength ($N \neq Z$) in units of $\mu_N^2$.}
\begin{tabular}{ccccccc}
Nucleus & Transition & $\sum B(M1)$ & $\sum B(M1)_l$ & $\sum B(M1)_{\sigma}$ & 
$\Delta$ & $\theta(^{\circ})$\\
\tableline
$^{10}Be$ & $T=1 \rightarrow T=1$ & 2.0930 & 0.1266 & 1.622 & 0.3444 & 67.64\\
          & $T=1 \rightarrow T=2$ & 0.0597 & 0.1508 & 0.0932 & -0.1843 & 
141.01\\
          & Total & 2.1527 & 0.2744 & 1.7152 & 0.1601 & 83.33\\
          &        &        &        &         & \\
$^{22}Ne$ & $T=1 \rightarrow T=1$ & 2.2796 & 0.3597 & 1.8063 & 0.1134 & 
85.90\\
          & $T=1 \rightarrow T=2$ & 0.3398 & 0.2732 & 0.1559 & -0.0893 &
102.4\\
          & Total & 2.6194 & 0.6329 & 1.9621 & 0.0243 & 88.70\\
\end{tabular}
\end{table}

\begin{table}
\caption{The dependence of the interference angle $\theta$ on the details 
of the interaction for $^{28}Si$.}
\begin{tabular}{ccccccc}
$x$ & $y$ & $\sum B(M1)$ & $\sum B(M1)_l$ & $\sum B(M1)_{\sigma}$ & 
$\Delta$ & $\theta(^{\circ})$\\
\tableline
0 & 0 & 2.838 & 2.465 & 0.372 & -0.001 & 90.03\\
  &   &       &       &       &        & \\
1 & 0 & 5.616 & 2.014 & 4.096 & -0.494 & 94.93\\
  &   &       &       &       &        & \\
0 & 1 & 3.112 & 2.502 & 0.607 &  0.003 & 89.93\\
  &   &       &       &       &        & \\
1 & 1 & 4.445 & 2.242 & 2.391 & -0.189 & 92.33\\
  &   &       &       &       &        & \\
1.5 & 0.5 & 7.870  & 1.619  & 7.209  & -0.988 & 98.31\\
\end{tabular}
\end{table}

\end{document}